\documentclass[12pt]{article}
\usepackage{amsmath}
\usepackage{graphicx,psfrag,epsf}
\usepackage{enumerate}
\usepackage{natbib}
\usepackage{subcaption}
\usepackage{url} 

\begin{document}

\def\spacingset#1{\renewcommand{\baselinestretch}
{#1}\small\normalsize} \spacingset{1}


\title{\bf Bayesian Data Augmentation for Partially Observed Stochastic Compartmental Models}
  \author{Shuying Wang\hspace{.2cm}\\
    Department of Statistics and Data Sciences, University of Texas at Austin\\
    and \\
    Stephen G. Walker\\
    Department of Statistics and Data Sciences, University of Texas at Austin}
    \date{}
    \maketitle

\begin{abstract}
Deterministic compartmental models are predominantly used in the modeling of infectious diseases, though stochastic models are considered more realistic, yet are complicated to estimate due to missing data. In this paper we present a novel algorithm for estimating the stochastic SIR/SEIR epidemic model within a Bayesian framework, which can be readily extended to more complex stochastic compartmental models. Specifically, based on the infinitesimal conditional independence properties of the model, we are able to find a proposal distribution for a Metropolis algorithm which is very close to the correct posterior distribution. As a consequence, rather than perform  a Metropolis step updating one missing data point at a time, as in the current benchmark Markov chain Monte Carlo (MCMC) algorithm, we are able to extend our proposal to the entire set of missing observations. This improves the MCMC methods dramatically and makes the stochastic models now a viable modeling option. A number of real data illustrations and the necessary mathematical theory supporting our results are presented.
\end{abstract}


\spacingset{1.5}
\section{Introduction}
\label{s:intro}
Compartmental models, at any given point in time, partition a population of individuals  into different compartments or states. The aim is to model the transitions of individuals between compartments. The deterministic compartmental models are predominantly used in analyzing outbreaks of infectious diseases, see for example \citet{Dehning2020} used the deterministic SIR model to fit the COVID-19 outbreak in Germany
assuming that there is random noise in the real-world observations but the underlying epidemic process is deterministic.
The deterministic models are usually simpler to handle because of the existence of analytical solutions to the differential equation system, but the stochastic SIR or SEIR models are considered more realistic due to the nature of the epidemic processes \citep{Roberts2015}. Further, stochastic models are more flexible when modeling the time-varying transition rates \citep{Roberts2015}. Also, the deterministic models are only appropriate for a sufficiently large population size, because the number of individuals are considered continuous variables in the differential equations while they are actually discrete \citep{Brauer2008}. Therefore, ``stochastic models remain preferable when their analysis is possible" \citep{ho2018}. However, the current benchmark Bayesian MCMC approaches using data augmentation with the partially observed epidemic processes \citep{gibson1998, oneill1999} are not up to the task, because they are ``very challenging and time-consuming, and for large systems with many hidden states, they can become computationally infeasible", which limits its use in practical applications
\citep{challenges2022}. Other methods include  Approximate Bayesian Computation (ABC) \citep{mckinley2009, blum2010, neal2012}, and sequential Monte Carlo (SMC) \citep{dukic2012, king2016} suffer from the same problem. In this paper, we develop a new MCMC algorithm with a novel data augmentation method that used the infinitesimal independence property of stochastic compartmental models to solve the missing data problem.

With the general stochastic epidemic SIR model \citep{bailey1975}, the population is divided into three compartments; $S=$ susceptible, $I=$ infected and $R=$ recovered, with two types of transition; from susceptible to infectious and from infectious to recovery. It is assumed not possible to return from recovered to either susceptible or infected. More complex epidemic models incorporate additional compartments and types of transition. For example, the SEIR model includes the incubation period of the disease in a compartment called ``exposed", labeled as $E$. 

The foundational framework of stochastic compartmental models lie in non--homogeneous Poisson processes which are assumed to have the Markov property. Interest focuses on the transition rates between compartments. For the SIR model, the parameters are the infection rate $\beta$ and the recovery rate $\gamma$, which yield the reproduction number $\beta/\gamma$ representing the expected number of new infections from a single infected individual from the population. The major difficulty for likelihood--based statistical inference  is that the epidemic process is often, if not always, partially observed; see \citet{bailey1975}. However, the likelihood function can only be tractable with the complete process; it being infeasible to integrate out the unobserved part of the process. There are two kinds of scenarios here: Firstly, in a given time period, the number of infections and recoveries are known, but the exact time points when these events occur are unknown. For example, there can be a daily report of diagnosed and recovered cases. Secondly, either the infection or recovery process can be completely unobserved, which means the number of infections or recoveries are also unknown. There can, for example,  be a daily report of the diagnosed cases but with no information about recoveries. In this case, the problems are more difficult to handle because the unobserved cases also comes with an unknown dimension.

Numerous methods have been developed to solve these two problems. Martingale methods and the EM algorithm have been used to obtain maximum likelihood estimators of the infection rate with a completely observed recovery process and a partially observed infection process; see \citet{becker1993, becker1997}. For Bayesian approaches using MCMC methods, sampling the unobserved part of the epidemic process as latent variables, and therefore obtaining the full likelihood and posterior distribution of the parameters when the infection or removal process is completely unobserved, has been described in \citet{gibson1998, oneill1999, gibson2001}. \citet{rose2020} adopted this MCMC data augmentation approach to fit a stochastic compartmental model for infestation data to help locate infested homes in urban areas. Another recent application of this method is by \citet{pooley2020}, who introduced a software tool called SIRE to estimate genetic and non-genetic effects in epidemic processes.
Further, \citet{cauchemez2008} used a diffusion process to approximate the number of infected in the SIR epidemic process for data augmentation, but this kind of approximation is less accurate with small populations.
\citet{keeling2008} proposed a computational method using the Kolmogorov forward equation, in which they were able to compute the likelihood value of the SIR model with incomplete data; however, the high computational demand can be a problem because this method involves the calculation of a matrix exponential. Another computational method reparameterizes the stochastic compartmental process into a multivariate birth process and estimates the transition probabilities using a Laplace transformation, which is more computationally efficient than the matrix exponential method; see \citet{crawford2012, crawford2018, ho2018}. However, these two computational methods are only applicable when both infection and removal processes are discretely observed, so they cannot deal with the situation when the number of infections or recoveries are unknown. 

This paper focuses on the classic MCMC algorithm proposed by \citet{oneill1999} who developed a proposal distribution for the missing observations. The limitation of the existing MCMC algorithm is the convergence speed; when sampling the unobserved infection process with a reversible jump algorithm, the step size of the Metropolis--Hastings sampler is small because only one time point is updated at each iteration, hence it is extremely slow to explore the parameter space, especially with large data sets \citep{challenges2022}. Our contribution is to introduce a new proposal idea which is developed to sample the unobserved infection or recovery process with larger step sizes at each iteration, and as we shall see, greatly improving the convergence speed, to such an extent that we believe the Bayesian MCMC could become the algorithm of choice for estimating SIR and SEIR models.  

Before proceeding, we briefly mention some further relevant papers. \citet{ball1995} showed that a general branching process; i.e. a linear birth--death process can be used to approximate the general epidemic process in the early stage for a large population. 
\citet{lekone2006} built a discrete-time SEIR model to approximate the underlying continuous-time process for an outbreak of Ebola by using binomial distributions to model the number of transitions in each time interval.
\cite{jewell2009} used an SIR type model for emerging infectious diseases in UK farms. They used a non--centered parameterization for the unobserved infection times and also used a reversible jump sampler to deal with the unknown number of infections.
\cite{dukic2012} fit a state--space extension of the discrete--time deterministic SEIR model with the Google flu trends data. They used a discretized time--step of one week and assumed that the epidemic process was observed with error.
\cite{neal2017} used a non--centered parameterization to construct a collapsing MCMC algorithm with the infection rate parameter integrated out when sampling other variables in the chain.
Finally, \citet{fintzi2021} used linear noise approximation to approximate the transition density of SIR process between discrete time points with a Gaussian density and implement it into a data augmentation MCMC framework.

The layout of the paper is as follows. In Section~\ref{s:method1} we review the current standard MCMC algorithm, and in Section~\ref{s:method2} we introduce our algorithm based on the infinitesimal conditional independence structure of the model. Section~\ref{s:math} details necessary mathematical background and Section~\ref{s:ill} contains illustrations and examples. Section~\ref{s:con} is a conclusion section and the Appendix provides a proof to one of the results established in Section~\ref{s:math}. R code for the illustrations and examples is provided in the GitHub link\footnote{https://github.com/ShuyingWang/SIR-SEIR-Model-MCMC}.

\section{Model and Benchmark MCMC Algorithm}
\label{s:method1}
\subsection{The SIR Model}
The general SIR model divides the population into three compartments, namely susceptible, infected, and recovered/removed. For notation we write $(S_t,\ I_t,\ R_t)$ to denote the number of individuals in each compartment at time $t\in[0, T]$. It is assumed that the population is closed without any natural births, deaths or immigration, and the recovered individuals are immune from further infection. Based on these two assumptions, there are two types of transition between the three compartments. Infection is the transition from susceptible to infected and recovery is the transition from infected to recovered. In this paper we look at the two transition processes, the infection process $X=\{X_t\}_{0\le t\le T}$, and the recovery process  $Y=\{Y_t\}_{0\le t\le T}$. Therefore, with the population size fixed at $N$ and $X_0=Y_0=0$, the state $(S_t,\ I_t,\ R_t)$ can be fully represented by $(X_t, Y_t)$ at a given time point $t$, with $S_t = S_0 - X_t$, $I_t = I_0 + X_t - Y_t$, and $R_t = N - S_t - I_t$.
The infection rate is denoted by $\beta$, and the recovery rate is denoted by $\gamma$. With the Markov assumption, the time intervals between transitions are exponentially distributed. In a small time interval $(t, t+dt)$, the probability of an infection is $\  \beta\, S_t \,I_t\,dt/N$, and the probability of a recovery is $\gamma \,I_t \,dt$. 
The infinitesimal probabilities are given by
\begin{equation*}
\begin{split}
     & P(X_{t+dt} = X_t+1\ |\ S_t,I_t)\ =\  \beta\,S_t\,I_t\,dt/N+o(dt) \quad\mbox{and} \\
     & P(Y_{t+dt} = Y_t+1\ |\ S_t,I_t)\ =\  \gamma I_t dt+o(dt),
\end{split}
\end{equation*}
with $t$ restricted to the time interval $[0, T]$. Denote the total number of infections and recoveries during this time period by $(n_x,n_y)$, and denote the time points for the occurrence of transitions  by $(\{t_i^x\}_{i=1}^{n_x},\ \{t_i^y\}_{i=1}^{n_y})$. 
Let $(x, y)$ denote the right continuous sample paths of $\{X_t, Y_t\}_{0\le t\le T}$, characterized by $(n_x, \{t_i^x\}_{i=1}^{n_x})$ and $(n_y, \{t_i^y\}_{i=1}^{n_y})$. The path functions are given by,
\begin{equation}
\label{path}
    x(t) = \sum_{i=1}^{n_x} \mathbf{1} (t_i^x \le t)\quad \mbox{and} \quad
    y(t) = \sum_{i=1}^{n_y} \mathbf{1} (t_i^y \le t),\ \ \ \forall\ t\in[0, T].
\end{equation}
Now let $S(t)$ and $I(t)$ denote the path functions of $S_t$ and $I_t$ represented by $x$ and $y$, and let
$S(t_i^x-) = S_0 - (i-1) \quad\mbox{and}\quad I(t_i^x-) = I_0 + (i-1) - Y_{t_i^x}$
be the state of $(S_t,\ I_t)$ just before the $i$th infection event, with $t_i^x-$ being the left limit of $t_i^x$, and similarly for $I(t_i^y-)$. 

Assume the initial state $(S_0, I_0, R_0)$ is known, then the likelihood function of the process $(X, Y)$ between times 0 and $T$ is given by,
\begin{equation}
\label{likelihood}
\begin{split}
    p(x, y\ |\ \beta,\gamma) 
    = \ & \text{exp}\left\{-\int_0^T\beta\frac{S(t)I(t)}Ndt\right\}\ \prod_{i=1}^{n_x}\ \beta\frac{S(t_i^x-)I(t_i^x-)}N \\
    & \text{exp}\left\{-\int_0^T\gamma I(t)dt\right\}\ \prod_{i=1}^{n_y}\ \gamma I(t_i^y-).
\end{split}
\end{equation}
Introducing conjugate Gamma priors $\pi_0(\beta) \equiv \text{Gamma}(a_\beta,\ b_\beta)$ and $\pi_0(\gamma) \equiv \text{Gamma}(a_\gamma,\  b_\gamma)$,
it is straightforward to sample from the full conditional posterior distributions of the parameters $(\beta, \gamma)$, given by
\begin{equation}
\label{conditionals}
    \begin{split}
        & p(\beta\ |\ -)\equiv \text{Gamma}\left(n_x+a_\beta,\ N^{-1}\int_0^TS(t)I(t)dt+b_\beta\right) \\
        & p(\gamma\ |\ -)\equiv \text{Gamma}\left(n_y+a_\gamma,\ \int_0^TI(t)dt+b_\gamma\right),
    \end{split}
\end{equation}
where the $-$ denotes all other variables. 
However, as has been previously mentioned, this simplicity with these distributions is only the case when the process is completely observed.

\subsection{Current Benchmark MCMC Algorithm}
Although stochastic compartmental models have a closed form likelihood function, it is hard to integrate out unobserved states when the data is incomplete. A standard Bayesian approach is to implement a data augmentation scheme by using an MCMC algorithm to sample the unobserved part of the process as latent variables; see \citet{gibson1998, oneill1999, gibson2001}. The problem setting is typically that one of the infection and recovery processes is completely unobserved and the other partially or completely observed. In the following we assume that the initial state is known, and there is a single observation of the infection process at time point $T$, and the recovery process is completely unobserved. The method described can easily be extended to multiple observations at different discrete time points.

Consequently, the observed data consists of the initial state of the process, i.e. $(S_0,\ I_0,\ R_0)$, and a single observation $X_T = n_x$. The unobserved latent variables include the time point of infections $(t_i^x)_{i=1}^{n_x}$, and the whole recovery process $(Y_t)_{0\le t\le T}$. The target is to estimate the parameters $(\beta, \gamma)$ as well as the number of recoveries $n_y$ during the time interval $[0, T]$.

With the conjugate Gamma priors, the parameters $(\beta, \gamma)$ can be directly sampled from their full conditional distributions via (\ref{conditionals}). It is also straightforward to sample the infection time points $(t_i^x)_{i=1}^{n_x}$ one at a time using a Metropolis--Hastings step within a Gibbs framework,
\begin{equation*}
 p(t_i^x\mid-)\propto\ S(t_i^x-)I(t_i^x-)\prod_{j=1}^{n_y}I(t_j^y-) \exp\left\{\int_0^T \beta\frac{S(t)I(t)}N+\gamma I(t) dt\right\}.
\end{equation*}
However, the latent recovery process $(Y_t)_{0\le t\le T}$ cannot be updated in this way because the number of time points $n_y$ is unknown. The classic MCMC algorithm samples this kind of latent process using a reversible jump component within the MCMC algorithm. There are three possible types of move at each iteration. A proposal to add one more time point, delete an existing time point, or move an existing time point. When adding a new time point, it is sampled uniformly in $[0,\ T]$. When removing a time point, it is uniformly selected from all the existing time points. When moving a time point, a combination of the above two steps is used. Succinctly, and with acceptance probabilities attached,
\begin{itemize}
        \item Add a new time point: $\ t'\sim U(0,T),\quad n_y' = n_y + 1,\quad \alpha(y,\ y')=\text{min}\left\{1,\  \frac{p(y'\mid-)T}{p(y\mid-)(n_y+1)}\right\}$.\par
        \smallskip
        \item Remove an existing time point: 
        $n_y' = n_y-1,\quad\alpha(y,\ y')=\text{min}\left\{1,\  \frac{p(y'\mid-)n_y}{p(y\mid-)T}\right\}$.\par
        \smallskip
        \item Move an existing time point: 
        $n_y' = n_y,\quad\alpha(y,\ y')=\text{min}\left\{1,\  \frac{p(y'\mid-)}{p(y\mid-)}\right\}.$
\end{itemize}
There are some limitations with the reversible jump part. The step size is very small, because only one change can occur at each iteration, so convergence and mixing can be slow. There is also high auto--correlation between iterations, which means a run of the algorithm requires a large number of iterations to get suitable posterior samples.

\section{A New MCMC Algorithm}
\label{s:method2}
\subsection{A Likelihood Factorization}
The Metropolis--Hastings procedure relies heavily on a good proposal distribution; as close to the target distribution as possible, while being able to sample easily from the proposal. 

The target distribution for sampling the recovery process $(Y_t)_{0\le t\le T}$ is given by
\begin{equation}
\label{target}
\begin{split}
p(y\mid-)\propto\  \prod_{i=1}^{n_x}\ I(t_i^x-)\   \text{exp}\left\{-\int_0^T\beta\frac{S(t)I(t)}N\,dt\right\}
\prod_{i=1}^{n_y}\ \gamma I(t_i^y-)\ \text{exp}\left\{-\int_0^T\gamma I(t)dt\right\}.
\end{split}
\end{equation}
Our aim is to find a proposal distribution which approximates this very well.

To this end, consider a time--inhomogeneous birth process $Y' = \{Y'_t\}_{0\le t\le T}$, with  intensity function $\lambda_k^y(t) = [\gamma(I_0 + x(t) - k)]^+ = \max\{0,\ \gamma(I_0 + x(t) - k)\}$, which is the birth rate for a birth process at time point $t$ when $Y'_t=k$. Here $x(t)$ is considered to be a deterministic function of $t$, as defined in (\ref{path}). With the number and time points of births denoted by $n_y$ and $\{t_i^y\}_{i=1}^{n_y}$, the likelihood function of this time--inhomogeneous birth process $Y'$ is given by
\begin{equation}
\label{fygivenx}
    f_{Y\mid X}(y\mid x) = \text{exp}\left\{-\int_0^T\gamma I(t)dt\right\}\ \prod_{i=1}^{n_y}\ \gamma I(t_i^y-),\ 
\end{equation}
where $I(t_i^y-) = I_0 + x(t_i^y) - (i-1)$. We will use this time-inhomogeneous birth process as the proposal distribution for process $Y$. Likewise, consider the time--inhomogeneous birth process $X'$ with intensity function 
$\lambda^x_k(t) = \left[\beta(S_0-k)(I_0 + k - y(t))/N\right]^+,$
and likelihood function
\begin{equation}
\label{fxgiveny}
    f_{X\mid Y}(x\mid y) = \text{exp}\left\{-\int_0^T\beta\frac{S(t)I(t)}N\,dt\right\}\ \prod_{i=1}^{n_x}\ \beta\frac{S(t_i^x-)I(t_i^x-)}N.\ 
\end{equation}
It is not surprising to find that the likelihood function of the general epidemic process in (\ref{likelihood}) can be factorized as the product of the likelihoods of these two time--inhomogeneous birth processes in (\ref{fygivenx}) and (\ref{fxgiveny}). For notation simplicity, for the following, we are going to write the likelihood using a factorization of the form 
\begin{equation}
\label{factor}
    p(x,\ y)=f_{X \mid Y}(x\mid y)f_{Y \mid X}(y\mid x),
\end{equation}
where $f_{X \mid Y}$ and $f_{Y \mid X}$ are given by (\ref{fxgiveny}) and (\ref{fygivenx}), respectively.
Note these are not the conditional distribution from $p(x,y)$.

This factorization property can be easily extended to other stochastic compartmental models. For a compartmental model with three types of transitions, such as the epidemic SEIR model, the likelihood can be expressed as the product of three time--inhomogeneous birth process densities.

\subsection{The New MCMC Algorithm}

For the new MCMC algorithm we introduce, the same approach as the current MCMC algorithm is used to update the parameters $\beta$, $\gamma$ and the partially observed infection process $X$ as in Section \ref{s:method1}, while a different approach is used to update the completely unobserved recovery process $Y$. The current MCMC algorithm uses a reversible jump method to update $Y$; the new MCMC algorithm takes advantage of the likelihood factorization (\ref{factor}) to develop a new proposal distribution for the $Y$ process.

It is not difficult to generate a time--inhomogeneous birth process when the intensity rate function is known, and it is even easier when the time--dependent term $x(t)$ is piece-wise constant. Therefore, the proposal distribution we use for the missing $Y$ process is $f_Y(y \mid x)$, i.e. (\ref{fygivenx}). The acceptance probability is given by
\begin{equation*}
\begin{split}
        \alpha(y,\ y')
        =\ \min\left\{1,\  \frac{f_X(x\mid y')f_Y(y'\mid x)f_Y(y\mid x)}{f_X(x\mid y)f_Y(y\mid x)f_Y(y'\mid x)} \right\}
        =\ \min\left\{1,\ \frac{f_X(x\mid y')}{f_X(x\mid y)} \right\},
\end{split}
\end{equation*}
where $Y'$ represents the proposed value.
With this proposal distribution, the whole recovery process $\{Y_t\}_{0\le t\le T}$ will be updated, so
the step size becomes large, while the acceptance rate remains acceptable.  
In fact, as will be shown, the proposal distribution is close to the target distribution; see Section \ref{s:math}.

Sampling from the time--inhomogeneous birth process is similar to sampling from a time--inhomogeneous Poisson process. The only difference between these two is that the intensity of the time--inhomogeneous birth process will depend on both time and the current state of the process, yet the intensity of a Poisson process will only depend on time. Since the current state of the process will stay unchanged between each jump, we can treat the birth process as a piece-wise Poisson process.

The time--inhomogeneous Poisson process can be simulated using a standard Poisson process; \citep{kingman1992}. To be more specific, for a time--inhomogeneous Poisson process with intensity function $\lambda(t)$, the integral of the intensity function is denoted by $\Lambda(t)=\int_0^t\lambda(s)ds$, which is the expected number of jumps between time 0 and $t$. The density of the waiting time before the first jump is given by $f_T(t)=\lambda(t) e^{-\Lambda(t)}$, and can be sampled from using the transformation of a standard exponential distribution. Take a standard exponential random variable sample $u \sim \text{exp}(1)$, then the waiting time before the first jump is given by $t = \Lambda^{-1}(u)$. The transformation of the distribution is given by
$f_T(t) = p(u)\ du/dt = \lambda(t) e^{-\Lambda(t)}.$
Similarly, for a time--inhomogeneous birth process $Y'$ with intensity function $\lambda_i^y(t)$, where $i$ represent the state of $Y'$ at time $t$, the waiting time before the next jump, namely the time interval between the $i$ th and $(i+1)$ th jump, can be sampled by $t_{i+1}^y - t_i^y = \Lambda_i^{-1}(u_i)$, where $\Lambda_i(t) = \int_{t_i^y}^{t_i^y+t}\lambda_i^y(s)\ ds$ and $u_i\sim\text{exp}(1)$.
Hence, the distribution of $t_{i+1}^y - t_i^y$ conditional on $t_i^y$ can be understood from
\begin{equation*}
\begin{split}
p(t_{i+1}^y-t_i^y\mid t_i^y) =  \frac{p(u_i)du_i}{d(t_{i+1}^y-t_i^y)} = \lambda_i^y(t_{i+1}^y) \,\exp\{-\Lambda_i(t_{i+1}^y-t_i^y)\}.
\end{split}
\end{equation*}
We start by sampling the first waiting time $t^y_1 = \Lambda_0^{-1}(u_0)$ with $u_0\sim \text{exp}(1)$, and then sample $t_{2}^y-t_1^y$ to get $t_2^y$. Keep sampling until $t_{i+1}^y > T$ and then take $n_y = i$. Therefore, the proposal sample $Y'$ can be written as a deterministic function of the i.i.d. standard exponential random variables $u=(u_0, u_1, \ldots, u_{n_y+1})$ and the proposal distribution can be written as
\begin{equation*}
\begin{split}
f_Y(y')  = &\ p(t_1^y)\ \ldots p(t^y_{n_y} -t^y_{n_y-1} \ |\ t^y_{n_y-1})\ P(t_{n_y+1}^y\ge T\ |\ t_{n_y}^y) \\
= & \prod_{i=0}^{n_y-1} \frac{p(u_i)du_i}{d(t_{i+1}^y-t_i^y)}P\left(u_{n_y}\ge \Lambda_{n_y}(T-t_{n_y}^y)\right) 
= p(u)\frac{\partial u}{\partial y'}.
\end{split}
\end{equation*}
This way of sampling a time-inhomogeneous birth process can also be used to adjust the step size of the proposal. The step size can be adjusted by only updating a randomly selected subset of $u$ at each iteration. The only problem is that the deterministic relationship between $u$ and $Y$ will be changed at each iteration since $(\beta,\ \gamma)$ and $X$ were also updated, so we cannot directly use the $u$ samples from the previous iteration. Instead, we need to recalculate $u$ at each iteration. Here we detail the algorithm:
\begin{enumerate}
\item[1.] Compute $u$ according to the current state of $(\beta,\ \gamma,\ X,\ Y)$, by the illustrated deterministic relationship between $Y$ and $u$.
\item[2.] Randomly select a subset of $u$. First, round $c(n_y+1)$ into an integer $m$, where $c$ represent the proportion of $u$ to be updated. Then, pick the index of the elements to be updated by uniformly sampling $m$ integers from $(0, 1, 2, \ldots, n_y)$.
Denote the subset to be updated by $u_\star$ and denote its compliment by $u_{-\star}$, so $u = (u_\star,\ u_{-\star})$.
\item[3.] Update $u_\star$ by sampling $u'_\star \stackrel{iid}\sim \text{exp}(1)$. The rest part of $u$ will stay the same, so set $u'_{-\star} = u_{-\star}$,\ \  $u' = (u_\star',\ u_{-\star}')$.
\item[4.] Compute $Y'$ using $u'$ and sample more $u'_\star \stackrel{iid}\sim \text{exp}(1)$ as needed when $n_y' > n_y$.
\end{enumerate}
In this way, the proposal distribution will become
$q(y') = p(u'_\star) \partial u'/\partial Y',$
and the acceptance probability will stay as 
$\alpha(y,\ y')=\text{min}\left\{1,\ f_X(x\mid y')/f_X(x\mid y) \right\},$
since
\begin{equation*}
        \frac{f_X(x\mid y')f_Y(y'\mid x)q(y)}{f_X(x\mid y)f_Y(y\mid x)q(y')} =
        \frac{f_X(x\mid y')p(u'_\star)p(u_{-\star})\frac{\partial u'}{\partial Y'}p(u_\star)\frac{\partial u}{\partial y}}{f_X(x\mid y)p(u_\star)p(u_{-\star})\frac{\partial u}{\partial y}p(u'_\star)\frac{\partial u'}{\partial y'}}
        = \frac{f_X(x\mid y')}{f_X(x\mid y)}.
\end{equation*}

With $u$ only partially updated, the step size will be smaller, so that the general acceptance rate can be improved. In this way, the acceptance rate can be adjusted to an ideal level by changing the size of $u_\star$.

\section{Mathematical Theory}
\label{s:math}
The aim in this section is to put down the details on the high acceptance probability for the proposal distribution.

The acceptance probability of the Metropolis--Hastings procedure can be understood as a distance measure between the target distribution in (\ref{target}) and proposal in (\ref{fygivenx}); 
\begin{equation*}
\begin{split}
d(p(y\mid x),\ f_Y(y\mid x))
=\ & 1-\int\int \min\{p(y'\mid x)f_Y(y\mid x),\ p(y\mid x)f_Y(y'\mid x)\}dydy'\\
=\ & 1-\int\int \min\left\{\frac{p(y'\mid x)f_Y(y\mid x)}{p(y\mid x)f_Y(y'\mid x)},1\right\}p(y\mid x)f_Y(y'\mid x)dydy'\\
=\ & 1-\text{E}[\alpha(Y,\ Y')].\\
\end{split}
\end{equation*}
This can be considered as an $\mathcal{L}_1$ distance, which can be upper bounded by the Hellinger distance \citep{kraft1955},
\begin{equation*}
\begin{split}
d(p(y\mid x),\ f_Y(y\mid x)) 
=\ & \frac12\int\int |p(y'\mid x)f_Y(y\mid x)-p(y\mid x)f_Y(y' \mid x)|dydy'\\
=\ & \frac12\ ||p(y'\mid x)f_Y(y\mid x),\ p(y\mid x)f_Y(y'\mid x)||_{\mathcal L_1}\\
\le\ & \sqrt2\ H(p(y'\mid x)f_Y(y\mid x),\ p(y\mid x)f_Y(y'\mid x)).\\
\end{split}
\end{equation*}
The Hellinger distance can be computed as
\begin{equation}
\label{hellinger}
    \begin{split}
         & H^2 (p(y'\mid x)f_Y(y\mid x),\ p(y\mid x)f_Y(y'\mid x)) \\
        =\ & 1-\int\int \sqrt{p(y'\mid x)f_Y(y\mid x)p(y\mid x)f_Y(y'\mid x)}dy\ dy' \\
        =\ & 1-\left\{\int f_Y(y\mid x)\sqrt{f_X(x\mid y)}dy\right\}^2/p(x) 
        =\ 1-\tilde p(x)^2/p(x),
    \end{split}
\end{equation}
where $p(x)=\int f_Y(y\mid x)f_X(x\mid y)dy\quad \text{and} \quad \tilde p(x)=\int f_Y(y\mid x)\sqrt{f_X(x\mid y)}dy$.
To illustrate clearly why $\tilde{p}(x)^2/p(x)$ is close to 1, we look at a simplified case. Consider a Poisson process $\{Y_t\}_{0\le t\le T}$ with intensity $\gamma$, and a time--inhomogeneous Poisson process $\{X_t\}_{0\le t\le T}$ that depends on $Y$, with intensity function $\beta(t) = \beta_0 + \epsilon Y_t$, so the infinitesimal probabilities of $(X,\ Y)$ are given by
\begin{equation*}
P(X_{t+h}=X_t+1\ |\ X_t,Y_t)=\beta(t)h+o(h), \quad\mbox{and} \quad
P(Y_{t+h}=Y_t+1\ |\ X_t,Y_t)=\gamma h +o(h).
\end{equation*}
In a small time interval $h$, $(X_{t+h},\ Y_{t+h})$ can be considered conditionally independent given the current state $(X_t,\ Y_t)$, i.e.
\begin{equation*}
       P(x(t+h),\ y(t+h)\ |\ x(t),\ y(t)) 
       = P(x(t+h)\ |\ x(t),\ y(t))\ P(y(t+h)\ |\ x(t),\ y(t)). 
\end{equation*}
The discrete--time analogue of a conditionally independent bivariate Markov chain $\{X_i,\ Y_i\}_{i=1}^n$ would have
\begin{equation*}
        P(x(n+1),\ y(n+1)\ |\ x(n),\ y(n))
        = P(x(n+1)\ |\ x(n),\ y(n))\ P(y(n+1)\ |\ x(n),\ y(n))
\end{equation*}
It is then clear that the joint distribution of $\{X_i,\ Y_i\}_{i=1}^n$ can be expressed as 
\begin{equation*}
\begin{split}
P(x,\ y)  = & \prod_{k=1}^n P(x(k),y(k)\ |\ x(k-1),y(k-1)) \\
 = &\prod_{k=1}^n P(x(k)\ |\ x(k-1),y(k-1)) P\big(y(k)\ |\ x(k-1),y(k-1)) 
 = P_X(x\ |\ y)\ P_Y(y\ |\ x),
\end{split}
\end{equation*}
where $P_X(x\ |\ y)=\prod_{k=1}^n P\big(x(k)\ |\ x(k-1),y(k-1)\big)$ itself can be considered as the distribution of a univariate time--inhomogeneous Markov process when $y$ is fixed; similarly for $P_Y(y\ |\ x)$.

Therefore, we can discretize the process by dividing the time interval $[0,\ T]$ into $n$ small time intervals of length $h = T/n$, and consider $(X_{t+h}-X_t,\  Y_{t+h}-Y_t)$ as two Bernoulli random variables that are independent to each other conditional on $(X_t, Y_t)$. This discretization enables us to perform the integrals involving $p(x)$ and $\tilde p(x)$ by summing over all the possible values of $\{Y_h, Y_{2h},\ldots, Y_{nh}\}$, and then let $n\rightarrow\infty$. When $X$ has no jumps between time 0 and $T$, we are able to compute the exact expression of the Hellinger distance ;
\begin{equation}
\label{Exactexp}
\begin{split}
         H^2(p(y'\mid x)f_Y(y\mid x),\ p(y\mid x)f_Y(y'\mid x)) =  1-\exp\left[-\gamma\left\{T-\epsilon^{-1}\left(e^{-\epsilon T}-4e^{-\frac12\epsilon T}+3\right)\right\}\right],
\end{split}
\end{equation}
which goes to $0$ as $\epsilon \rightarrow 0$. See the Appendix A of the Supplementary Materials for the mathematical details.

Returning to the likelihood factorization in (\ref{factor}), when we sample $Y$ from $f_Y(y\mid x)$, we are only using the information in $f_Y(y\mid x)$, and ignore the information in $f_X(x|y)$. However, this ignored information is negligible when the value of $\epsilon$ is small and recall that $\epsilon$ is the change in the intensity of the $X$ process when $Y$ changes by a single unit. Hence, our proposal is close to the target when the intensity of $X$ remains almost constant over a change of a single unit in $Y$.

\section{Illustrations}
\label{s:ill}
\subsection{A First Illustration using Simulated Data}
We start with a simple illustration where we assume that the recovery process $Y$ is completely observed and the parameters $\beta, \gamma$ are fixed and known, so only the sampling of the the unobserved infection process $X$ is required. We evaluated and compared the performance of the current and new MCMC algorithm by looking at the effective sample size of the random variable $n_x$, i.e. the total number of infections.

The population size was taken to be $N = 1,000,000$, with $I_0 = 100$ initial infections and $R_0 = 0$ initial recoveries. The epidemic process was simulated with parameters $\beta = 0.2$ and $\gamma = 0.2$ between time $0$ and $T = 10$; this specifically yielded $n_x = 195$ infections and $n_y=194$ recoveries. In both MCMC algorithms, i.e. ours and \cite{oneill1999}, the infection process $X$ was initialized by uniformly sampling $n_x$ from $100$ to $300$, with sampling each $t_i^x\sim U(0, 10)$. 

\begin{figure}
\centering\includegraphics[width=1\linewidth, height=10cm]{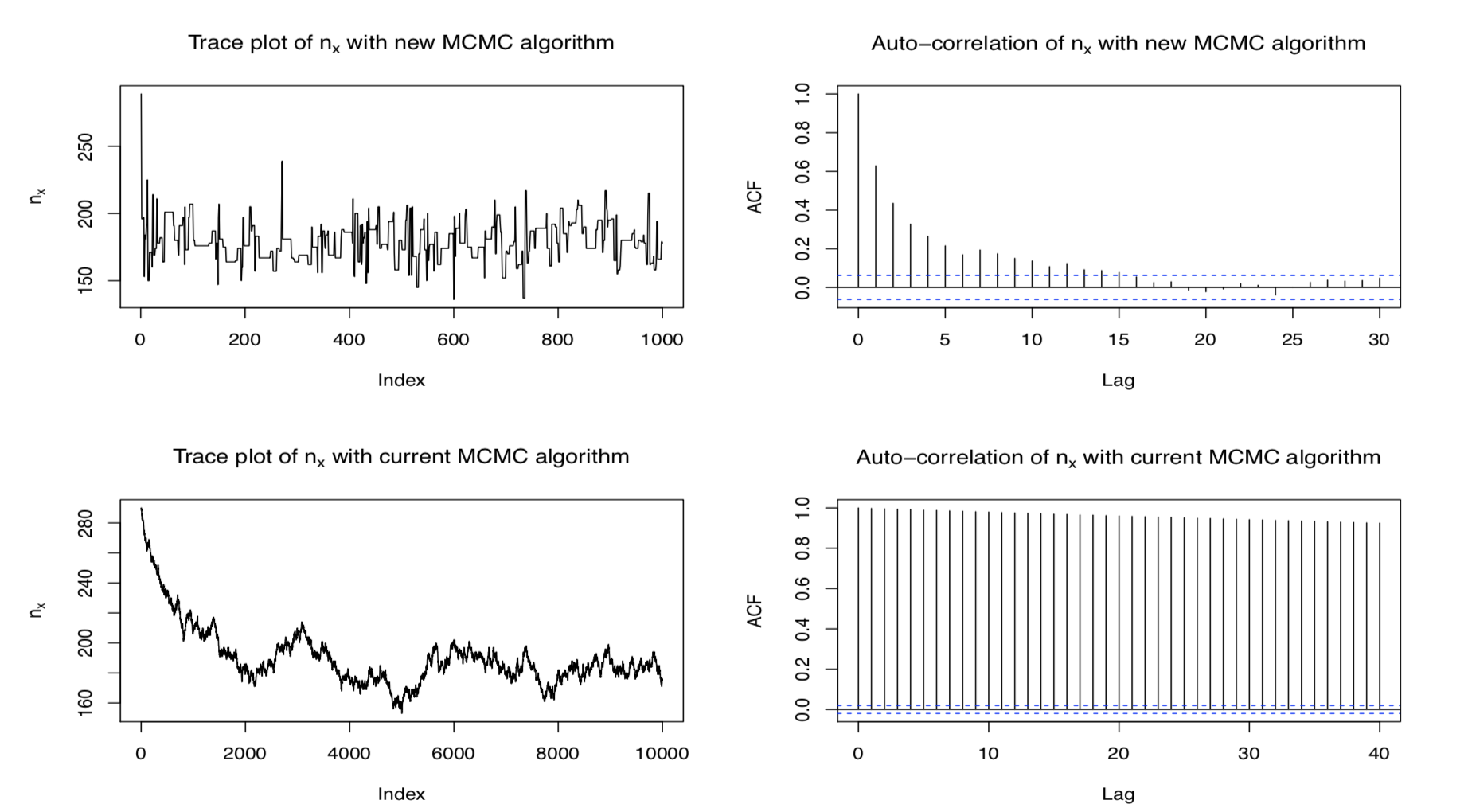}
\caption{Comparison for the first illustration using simulated data}
\label{fig1}
\end{figure}

For our MCMC algorithm, it took less than 5 seconds to get 1000 samples from the posterior, with an acceptance rate of 0.245, and an effective sample size of 119. This was computed using the R package ``mcmcse'' by \citet{mcmcse}. A trace plot of the output, demonstrating a very quick convergence is presented in Figure~\ref{fig1}. The auto-correlation was low, with the samples almost independent at lag 10; see Figure~\ref{fig1}.

On the other hand, for the \cite{oneill1999} MCMC algorithm, with which we are making a comparison, it took more than 1 minute to get 10,000 samples, with an acceptance rate of 0.936. However, the effective sample size was only 8 and convergence was slow, around 2000 iterations, see Figure~\ref{fig1}. Although the acceptance rate was high, a consequence of the small proposals, it is not surprising that the auto-correlation between iterations was also high because of the small step size at each iteration; see Figure~\ref{fig1}. See the Supplementary Material for the relevant R code.

\subsection{A Second Illustration using Simulated Data}

In the second illustration, we work on the opposite problem by assuming that the recovery process $Y$ is completely unobserved and the infection process $X$ is completely observed, with the initial states $(S_0, I_0, R_0)$ known. At the same time, we assume that the parameters $(\beta, \gamma)$ are unknown. 

The population size was taken as  $N = 1,000,000$, with $I_0 = 100$ initial infections and $R_0 = 0$ initial recoveries. The epidemic process was simulated with parameters $\beta = 0.25$ and $\gamma = 0.15$ between time $0$ and $T = 10$, yielding $n_x = 353$ infections and $n_y=231$ recoveries. 
We used the priors with parameters $a_\beta = b_\beta = a_\gamma = b_\gamma = 0.1$ and initialized the unobserved recovery process $Y$ by uniformly sampling $n_y$ from $200$ to $450$ and sampling each $t_i^y\sim U(0, 10)$. 

\begin{figure}
\centering\includegraphics[width=1\linewidth, height=10cm]{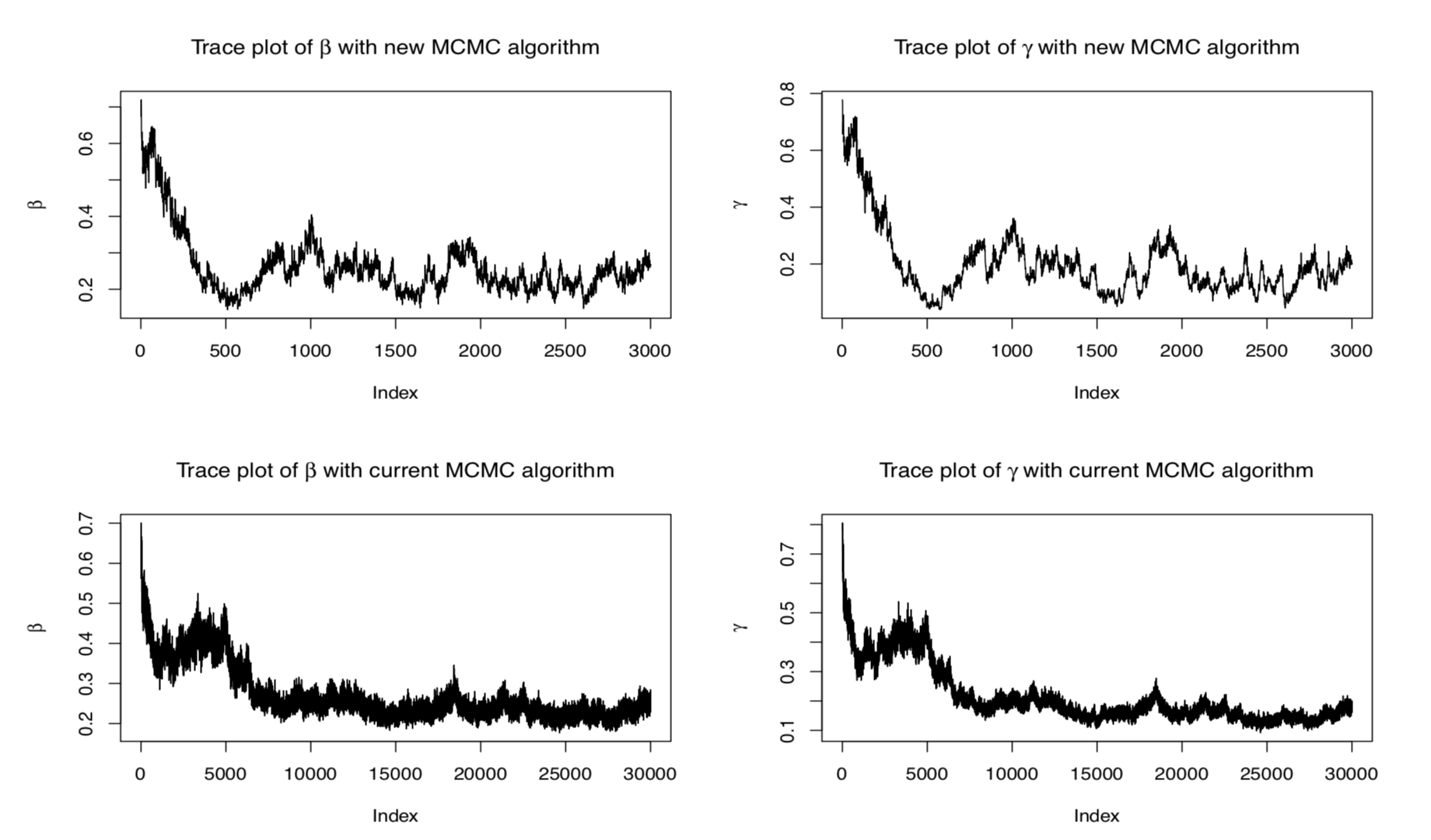}
\caption{Trace plots for the second illustration using simulated data}
\label{fig2}
\end{figure}

For our  MCMC algorithm it took around 30 seconds to run 3000 iterations. It took about 300 iterations to converge and after a burn in at the 300th iteration, we obtained the posterior sample means $\bar\beta=0.238$ and $\bar\gamma=0.165$. For the corresponding trace plots, see Figures~\ref{fig2}.

In comparison, when using the \cite{oneill1999} MCMC algorithm, it took around 450 seconds to run 30,000 iterations. It took about 8000 iterations to converge and after a burn in at the 8000th iteration, we obtained the posterior sample means as $\bar\beta=0.237$ and $\gamma=0.165$. For the trace plots, see Figures~\ref{fig2}. We can see the convergence speed of the \cite{oneill1999} MCMC algorithm is much slower than our MCMC algorithm. See the Supplementary Material for R code.

\subsection{SIR Model and Smallpox Data}
Our illustration with real data for the SIR model comes from the smallpox cases reported in Abakaliki, Nigeria, in 1967; see \cite{thompson1968}. The dataset is given by the removal times of 32 smallpox cases in a closed community with 120 individuals. \cite{oneill1999} have illustrated their algorithm with the same dataset, assuming the total number of infections was known. We were able to replicate their results with our MCMC algorithm. However, to better illustrate our method, we assume that the infection process was completely unobserved with unknown total number of infections.

\begin{figure}
\centering\includegraphics[width=1\linewidth, height=10cm]{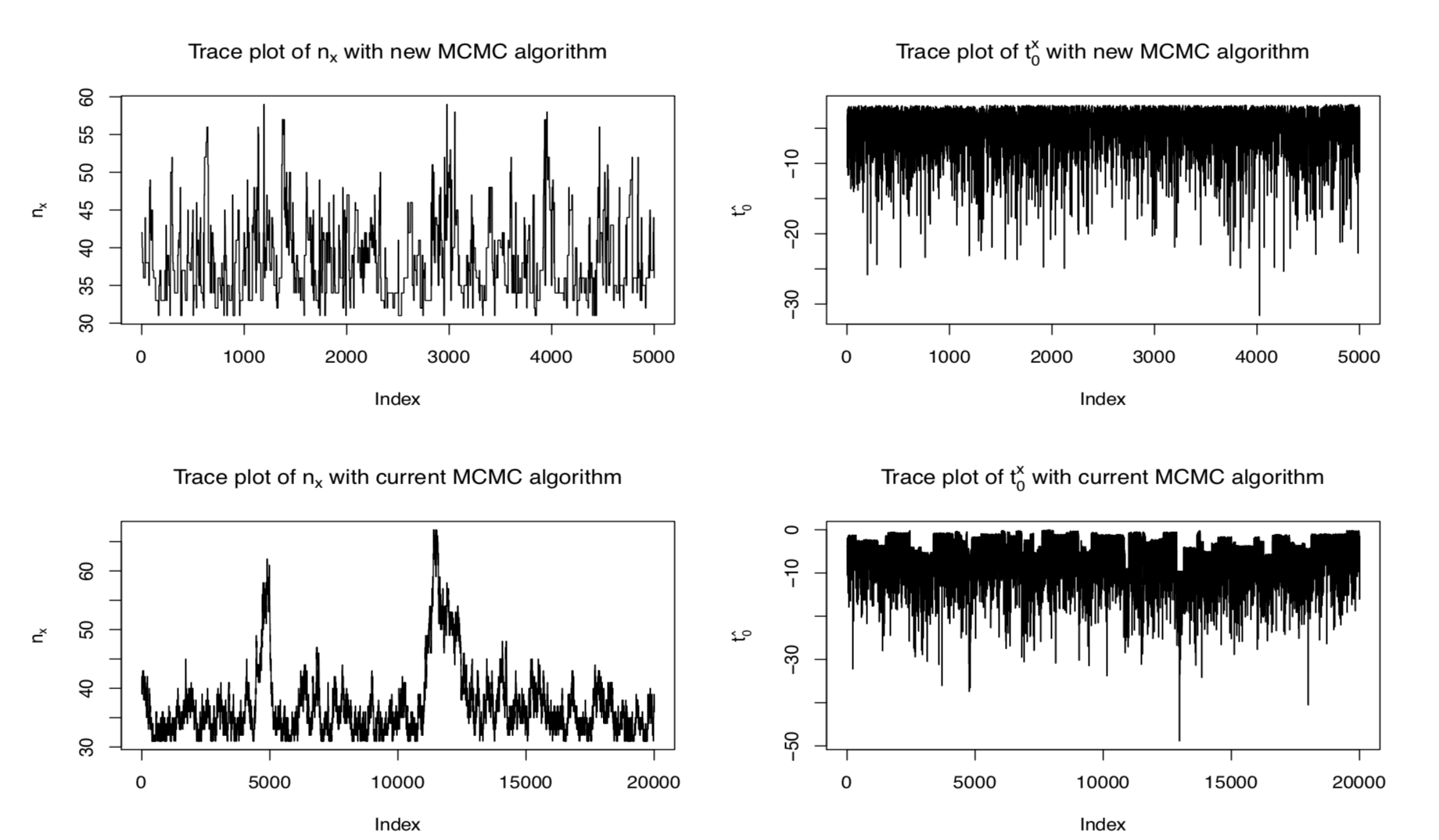}
\caption{Trace plots for the smallpox data illustration}
\label{fig3}
\end{figure}

\begin{figure}
\centering\includegraphics[width=1\linewidth, height=10cm]{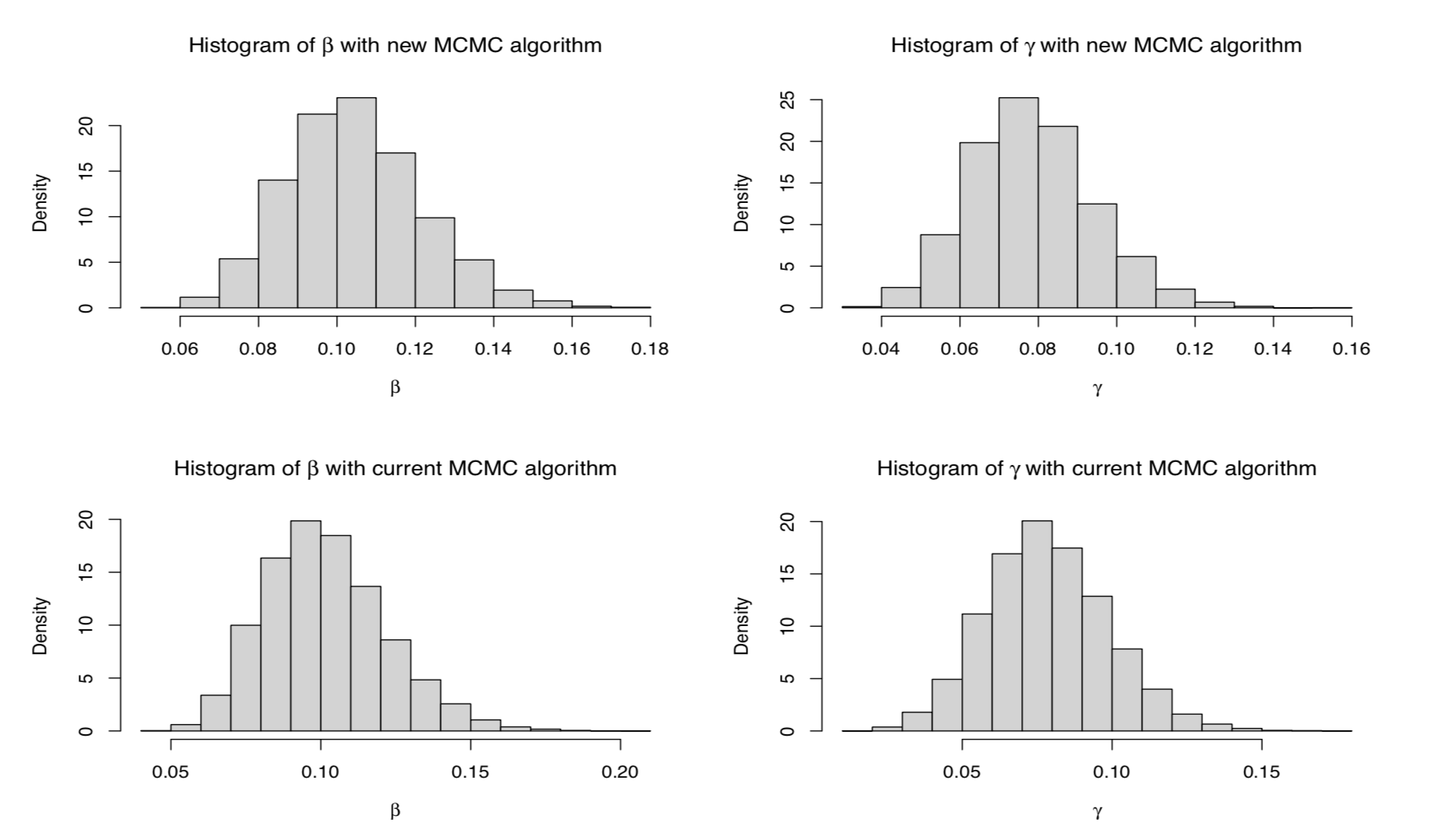}
\caption{Histograms for the smallpox data illustration}
\label{fig4}
\end{figure}

Different from the previous two illustrations with the simulated datasets, for the real smallpox dataset, we need to sample an extra initial infection time $t_0^x$, which is negative when the first removal time is indexed at 0. With prior $\pi_0(t_0^x) = \theta_0\exp(\theta_0 t_0^x)\mathbf{1}(t_0^x<0)$, the full conditional posterior of $t_0^x$ is given by
$p(t_0^x\mid-) = \theta e^{\theta( t_0^x-t_1^x)}\mathbf{1}(t_0^x<t_1^x),\quad\theta = \theta_0+\gamma+\beta S_0/N$.
We took the prior hyper-parameters $a_\beta = 10,\ b_\beta=100,\ a_\gamma=10, b_\gamma = 100$ and $\theta_0=0.1$. We used the method elaborated in Section~\ref{s:method2} to adjust the step size of our MCMC algorithm by only updating a subset of $u$ at each iteration to reach a moderate acceptance rate of 0.15. It only took 7 seconds to run 5000 iterations with the new MCMC algorithm. The posterior sample means and variances of the parameters are given by $\bar\beta=0.105$, and $\text{var}(\beta)=0.0003$, with $\bar\gamma=0.078$ and $\text{var}(\gamma) = 0.0002$. For associated trace plots and histograms, see Figures~\ref{fig3} and Figures~\ref{fig4}.

In comparison with the \cite{oneill1999} MCMC, to achieve a similar effective sample size, the \cite{oneill1999} algorithm needed to run for 20,000 iterations, taking around 30 seconds, with an acceptance rate of 0.59. The posterior sample means and variances of the parameters are given by $\bar\beta=0.102$, $\text{var}(\beta)=0.0004$, and $\bar\gamma=0.078$ and  $\text{var}(\gamma) = 0.0004$. For corresponding trace plots and histograms, see Figures~\ref{fig3} and Figures~\ref{fig4}. See the Supplementary Material for R code.

\subsection{SEIR Model and COVID-19 Data}

The SEIR model divides the population into four compartments; $S=$ susceptible, $E=$ exposed, $I=$ infected and $R=$ recovered, with three types of transition; from susceptible to exposed, denoted by $X$, from exposed to infectious, denoted by $Y$ and from infectious to recovery, denoted by $Z$. The exposed compartment represents the incubation period of the disease and will eventually become infectious. Similar to the SIR model, it is assumed not possible to return from recovered to either susceptible or infected, and the recovery process actually includes both recoveries and deaths. Except for the infection rate $\beta$ and recovery rate $\gamma$, another parameter is introduced in the SEIR model, which is the transition rate from exposed to infectious, denoted by $\alpha$.
The likelihood of the process between time $0$ and $T$ is given by
\begin{equation*}
\begin{split}
    p(x, y, z\mid \beta, \alpha, \gamma)
    = \ & \text{exp}\left\{-\int_0^T\beta\frac{S(t)I(t)}Ndt\right\}\ \prod_{i=1}^{n_x}\ \beta\frac{S(t_i^x-)I(t_i^x-)}N \\
    & \text{exp}\left\{-\int_0^T\alpha E(t)dt\right\}\ \prod_{i=1}^{n_y}\ \alpha E(t_i^y-) \ 
    \text{exp}\left\{-\int_0^T\gamma I(t)dt\right\}\ \prod_{i=1}^{n_z}\ \gamma I(t_i^z-). 
\end{split}
\end{equation*}
As elaborated in Section \ref{s:method2}, a likelihood factorization is given by
$$p(x,\ y,\ z)=f_{X\mid Y,Z}(x\mid y,z)f_{Y\mid X,Z}(y\mid x,z) f_{Z\mid X,Y}(z\mid x,y),$$
where the three terms are conditional distributions but not those from the joint $p(x,y,z)$, though are close to them.
The second real data illustration comes from the COVID-19 cases reported in french overseas department Mayotte in March and April, 2020. The dataset is well documented in \citet{Manou2020}, given by the daily reports of confirmed and removed cases from March 13 to April 17 2020 in Mayotte, so the $Y$ and $Z$ processes were discretely observed, while the $X$ process is completely unobserved. We set March 13 as $t_0$ and April 17 as $T$, and used the same initial states as \citet{Manou2020} with $I_0 = 3$ and $E_0 = 15$.
A control measure was introduced in March 29 2020, so we considered day 16 as a change point for the infection rate $\beta$ and modeled two different values of the parameter $\beta = (\beta_1,\ \beta_2)$, for the two different stages of the epidemic process, i.e. before and after March 29.
\citet{Manou2020} used a deterministic SEIR model and estimated the infection rate $\beta$ with the other two transition rates fixed. We used a stochastic SEIR model and estimated all three transition rates. According to the prior information given in \citet{Manou2020}, the latent period is around 6 days and the infectious period is around 10 days in average, so we took informative priors for $\alpha$ and $\gamma$ with $a_\alpha = 100,\ b_\alpha=600,\ a_\alpha=50, b_\alpha = 500$, and weak priors for $\beta = (\beta_1,\ \beta_2)$ with $a_\beta=5,\ b_\beta=50$.

\begin{figure}
\centering\includegraphics[width=1\linewidth, height=10cm]{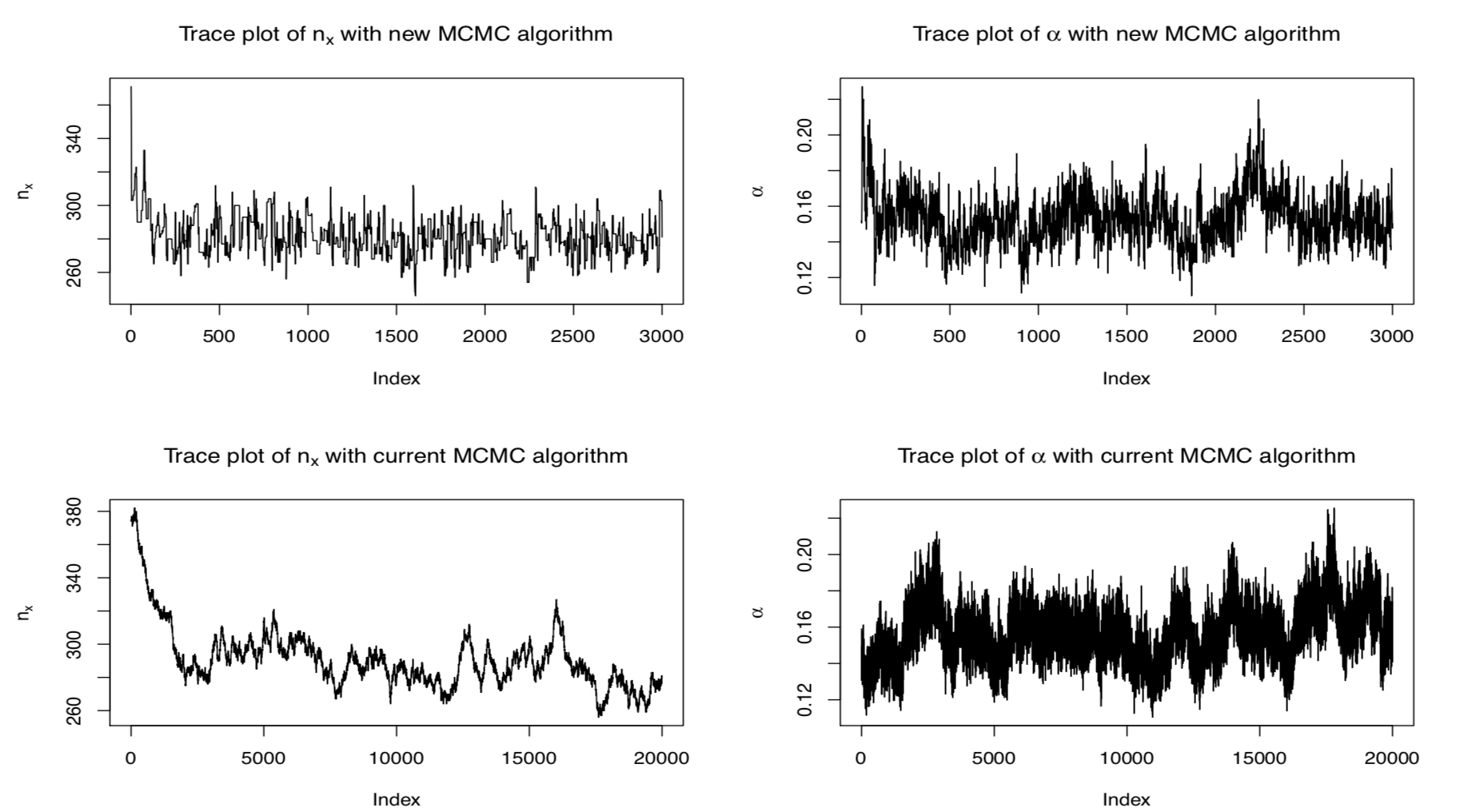}
\caption{Trace plots of $n_x$ and $\alpha$ for the COVID-19 data illustration}
\label{fig5}
\end{figure}

\begin{figure}
\centering\includegraphics[width=1\linewidth, height=10cm]{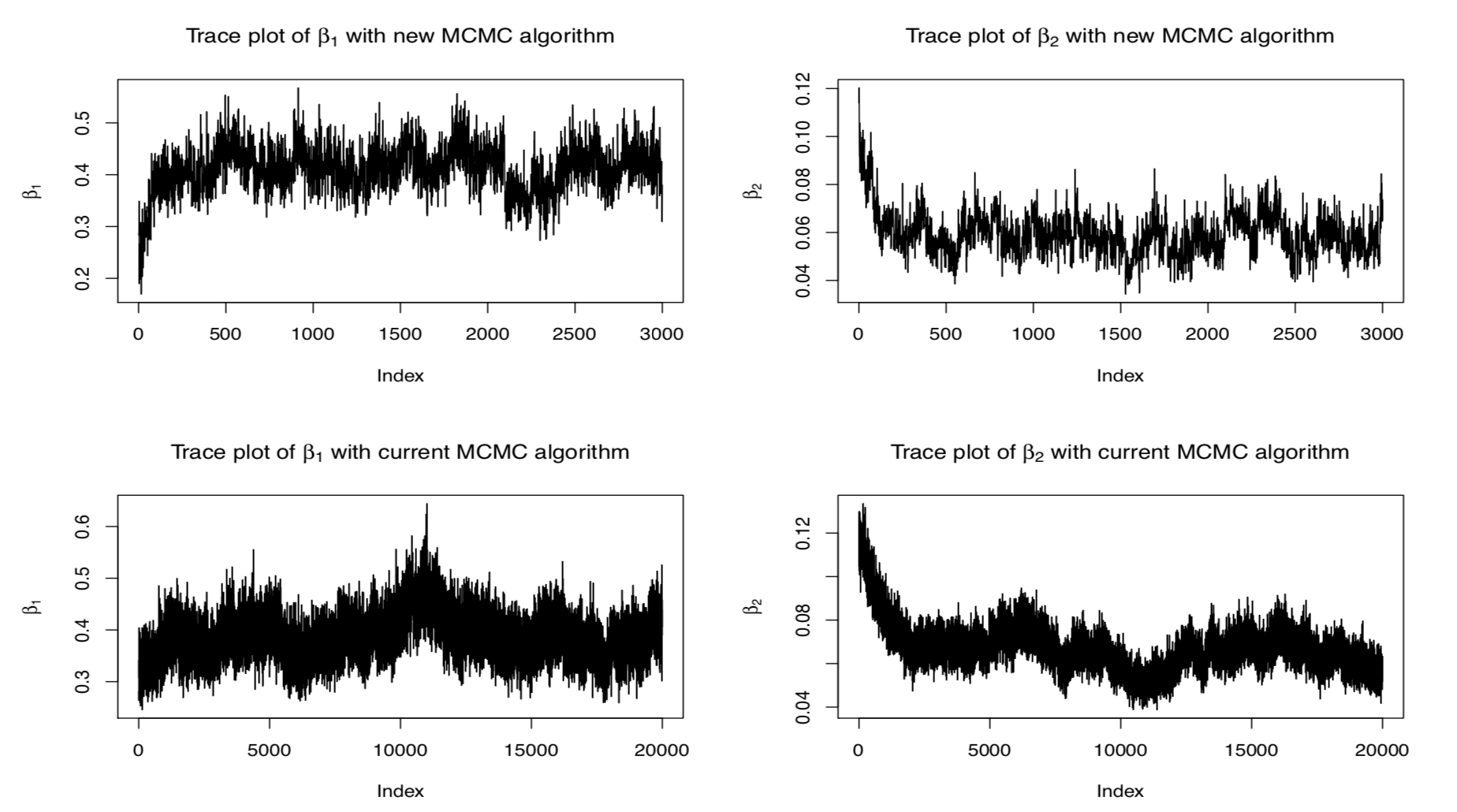}
\caption{Trace plots of $\beta_1$ and $\beta_2$ for the COVID-19 data illustration}
\label{fig6}
\end{figure}

For our MCMC algorithm it took 3000 iterations to get a well mixed posterior sample, which converges within 200 iterations. The posterior sample means of the parameters are given by $\bar\beta_1=0.402,\ \bar\beta_2=0.058,\ \bar\alpha = 0.152$ and $\bar\gamma = 0.054$.
In comparison, the \cite{oneill1999}  algorithm needed to run for 20,000 iterations to get a well mixed posterior sample and it took at least 2000 iterations to converge.
The posterior sample means of the parameters are given by $\bar\beta_1=0.388,\ \bar\beta_2=0.064,\ \bar\alpha = 0.157$ and $\bar\gamma = 0.054$. For the associated trace plots, see Figures~\ref{fig5} and Figures~\ref{fig6}. 

The results were verified by the R0 package \citep{Obadia2012} with the incidences data and the distribution of the generation time as inputs. We used 11 days as the mean generation time which is computed from the mean latent period plus a half of the mean infectious period \citep{svensson2007} 
with the prior information given in \citet{Manou2020}.
By using the R0 package, we got the Basic reproduction number $r_0 = 7.03$ with the exponential growth method, and $r_0 = 8.64$ with the maximum likelihood method before the change point, compared to $r_0 = \beta_1/\gamma=7.44$ with our posterior means. After the change point, we got $r_0 = 0.79$ with the exponential growth method, and $r_0 = 1.97$ with the maximum likelihood method, compared to $r_0 = \beta_2/\gamma=1.07$ with our posterior means. For all three methods, we can see a big drop of the basic reproduction number through the change point, but the $r_0$ computed by R0 package before the change point is closer to our results than that after the change point, which can be explained by the bias caused by the missing information of the initial value of $I$, since the incidences data alone cannot imply the current number of infectious individuals at any time point. See the Supplementary Material for R code.

\section{Conclusions}
\label{s:con}
In this paper we have developed and demonstrated a new Bayesian MCMC algorithm estimating compartmental models. In particular, we have provided a strategy for sampling a completely unobserved infection or recovery process in the general SIR/SEIR model. It is important to note that our method can easily be extended to any kind of stochastic compartmental model.

Current samplers for the Bayesian MCMC framework are highly problematic. A Metropolis--Hastings algorithm is more efficient when the proposal distribution is close to the target distribution and we have found a proposal that approximates the target distribution very well using a time-inhomogeneous birth process. Compared to current MCMC algorithms, which use a reversible jump procedure, using the time-inhomogeneous birth process as a proposal is advantageous because it does not depend on the sample from the previous iteration. Therefore, the auto-correlation between iterations will be low and the convergence speed will be fast. This class of proposal distribution now makes Bayesian MCMC analysis of stochastic compartmental models completely viable.

\bibliographystyle{apalike}
\bibliography{Bib}

\begin{thebibliography}{}

\bibitem[Bailey, 1975]{bailey1975}
Bailey, N. T.~J. (1975).
\newblock {\em Mathematical Theory of Infectious Diseases and Its Application}.
\newblock Griffin.

\bibitem[Ball and Donnelly, 1995]{ball1995}
Ball, F. and Donnelly, P. (1995).
\newblock Strong approximation for epidemic models.
\newblock {\em Stochastic Processes and their Applications}, 55:1--21.

\bibitem[Becker, 1993]{becker1993}
Becker, N.~G. (1993).
\newblock Parametric inference for epidemic models.
\newblock {\em Mathematical Biosciences}, 117:239--251.

\bibitem[Becker and Hasofer, 1997]{becker1997}
Becker, N.~G. and Hasofer, A.~M. (1997).
\newblock Estimation in epidemics with incomplete observations.
\newblock {\em Journal of the Royal Statistical Society}, 59(2):415--429.

\bibitem[Blum and Tran, 2010]{blum2010}
Blum, M. G.~B. and Tran, V.~C. (2010).
\newblock H{IV} with contact tracing: a case study in approximate {B}ayesian
  computation.
\newblock {\em Biostatistics}, 11(4):644--660.

\bibitem[Brauer, 2008]{Brauer2008}
Brauer, F. (2008).
\newblock Compartmental models in epidemiology.
\newblock {\em Mathematical Epidemiology}, 1945:19--79.

\bibitem[Cauchemez and Ferguson, 2008]{cauchemez2008}
Cauchemez, S. and Ferguson, N.~M. (2008).
\newblock Likelihood-based estimation of continuous-time epidemic models from
  time-series data: application to measles transmission in london.
\newblock {\em Journal of the Royal Society Interface}, 5:885--897.

\bibitem[Crawford et~al., 2018]{crawford2018}
Crawford, F.~W., Ho, L. S.~T., and Suchard, M.~A. (2018).
\newblock Computational methods for birth-death process.
\newblock {\em Wiley Interdisciplinary Reviews: Computational Statistics},
  10:1423.

\bibitem[Crawford and Suchard, 2012]{crawford2012}
Crawford, F.~W. and Suchard, M.~A. (2012).
\newblock Transition probabilities for general birth-death process with
  applications in ecology, genetics, and evolution.
\newblock {\em Journal of Mathematical Biology}, 65:553--580.

\bibitem[Dehning et~al., 2020]{Dehning2020}
Dehning, J., Zierenberg, J., Spitzner, F.~P., Wibral, M., Neto, J.~P., Wilczek,
  M., and Priesemann, V. (2020).
\newblock Inferring change points in the spread of {COVID}-19 reveals the
  effectiveness of interventions.
\newblock {\em Science}.

\bibitem[Dukic et~al., 2012]{dukic2012}
Dukic, V., Lopes, H.~F., and Polson, N.~G. (2012).
\newblock Tracking epidemics with {G}oogle flu trends data and a state-space
  {SEIR} model.
\newblock {\em Journal of the American Statistical Association},
  107(500):1410--1426.

\bibitem[Fintzi et~al., 2022]{fintzi2021}
Fintzi, J., Wakefield, J., and Minin, V.~N. (2022).
\newblock A linear noise approximation for stochastic epidemic models fit to
  partially observed incidence counts.
\newblock {\em To appear in Biometrics}.

\bibitem[Flegal et~al., 2021]{mcmcse}
Flegal, J.~M., Hughes, J., Vats, D., Dai, N., Gupta, K., and Maji, U. (2021).
\newblock {\em mcmcse: Monte Carlo Standard Errors for MCMC}.
\newblock Riverside, CA, and Kanpur, India.
\newblock R package version 1.5-0.

\bibitem[Gibson and Renshaw, 1998]{gibson1998}
Gibson, G.~J. and Renshaw, E. (1998).
\newblock Estimating parameters in stochastic compartmental models using
  {M}arkov chain methods.
\newblock {\em IMA Journal of Mathematics Applied in Medicine \& Biology},
  15:19--40.

\bibitem[Gibson and Renshaw, 2001]{gibson2001}
Gibson, G.~J. and Renshaw, E. (2001).
\newblock Likelihood estimation for stochastic compartmental models using
  markov chain methods.
\newblock {\em Statistics and Computing}, 11:347--358.

\bibitem[Ho et~al., 2018]{ho2018}
Ho, L. S.~T., Crawford, F.~W., and Suchard, M.~A. (2018).
\newblock Direct likelihood-based inference for discretely observed stochastic
  compartmental models of infectious disease.
\newblock {\em The Annals of Applied Statistics}, 12(3):1993--2021.

\bibitem[Jewell et~al., 2009]{jewell2009}
Jewell, C.~P., Kypraios, T., Neal, P., and Roberts, G.~O. (2009).
\newblock Bayesian analysis for emerging infectious diseases.
\newblock {\em Bayesian Analysis}, 4(4):465--496.

\bibitem[Keeling and Ross, 2008]{keeling2008}
Keeling, M.~J. and Ross, J.~V. (2008).
\newblock On methods for studying stochastic disease dynamics.
\newblock {\em Journal of the Royal Society Interface}, 5:171--181.

\bibitem[King et~al., 2016]{king2016}
King, A.~A., Nguyen, D., and Ionides, E.~L. (2016).
\newblock Statistical inference for partially observed {M}arkov process via the
  {R} package ``{P}omp".
\newblock {\em Journal of Statistical Software}, 69:1--43.

\bibitem[Kingman, 1992]{kingman1992}
Kingman, J. (1992).
\newblock {\em Poisson Processes}.
\newblock Oxford Studies in Probability. Clarendon Press.

\bibitem[Kraft, 1955]{kraft1955}
Kraft, C.~H. (1955).
\newblock Some conditions for consistency and uniform consistency of
  statistical procedures.
\newblock {\em University of California Publications in Statistics, Vol. 2}.

\bibitem[Lekone and Finkenst{\"a}dt, 2006]{lekone2006}
Lekone, P.~E. and Finkenst{\"a}dt, B. (2006).
\newblock Statistical inference in a stochastic epidemic {SEIR} model with
  control intervention: {E}bola as a case study.
\newblock {\em Biometrics}, 62:1170--1177.

\bibitem[Manou-Abi and Balicchi, 2020]{Manou2020}
Manou-Abi, S.~M. and Balicchi, J. (2020).
\newblock Analysis of the {COVID}-19 epidemic in french overseas department
  {M}ayotte based on a modified deterministic and stochastic {SEIR} model.
\newblock {\em medRxiv}.

\bibitem[McKinley et~al., 2009]{mckinley2009}
McKinley, T., Cook, A.~R., and Deardon, R. (2009).
\newblock Inference in epidemic models without likelihoods.
\newblock {\em The International Journal of Biostatistics}, 5(1):24.

\bibitem[Neal, 2012]{neal2012}
Neal, P. (2012).
\newblock Efficient likelihood-free {B}ayesian computation for household
  epidemics.
\newblock {\em Statistics and Computing}, 22:1239--1256.

\bibitem[Neal and Xiang, 2017]{neal2017}
Neal, P. and Xiang, F. (2017).
\newblock Collapsing of non-centred parameterized {MCMC} algorithms with
  applications to epidemic models.
\newblock {\em Scandinavian Journal of Statistics}, 44:81--96.

\bibitem[Obadia et~al., 2012]{Obadia2012}
Obadia, T., Haneef, R., and B{\"o}elle, P.-Y. (2012).
\newblock The r0 package: a toolbox to estimate reproduction numbers for
  epidemic outbreaks.
\newblock {\em BMC Medical Informatics and Decision Making}, 12:147--147.

\bibitem[O'Neill and Roberts, 1999]{oneill1999}
O'Neill, P.~D. and Roberts, G.~O. (1999).
\newblock Bayesian inference for partially observed stochastic epidemics.
\newblock {\em Journal of the Royal Statistical Society}, 162(1):121--129.

\bibitem[Pooley et~al., 2020]{pooley2020}
Pooley, C.~M., Marion, G., Bishop, S.~C., Bailey, R.~I., and Doeschl-Wilson,
  A.~B. (2020).
\newblock Estimating individuals’ genetic and non-genetic effects underlying
  infectious disease transmission from temporal epidemic data.
\newblock {\em PLoS Computational Biology}, 16:1008447.

\bibitem[Roberts et~al., 2015]{Roberts2015}
Roberts, M.~G., Andreasen, V., Lloyd, A.~L., and Pellis, L. (2015).
\newblock Nine challenges for deterministic epidemic models.
\newblock {\em Epidemics}, 10:49--53.

\bibitem[Rose et~al., 2020]{rose2020}
Rose, E.~B., Roy, J., Castillo-Neyra, R., Ross, M.~E., Condori-Pino, C.,
  Peterson, J.~K., N{\'a}quira-Velarde, C., and Levy, M.~Z. (2020).
\newblock A real-time search strategy for finding urban disease vector
  infestations.
\newblock {\em Epidemiologic Methods}, 9:20200001.

\bibitem[Svensson, 2007]{svensson2007}
Svensson, {\AA}. (2007).
\newblock A note on generation times in epidemic models.
\newblock {\em Mathematical biosciences}, 208(1):300--311.

\bibitem[Swallow et~al., 2022]{challenges2022}
Swallow, B., Birrell, P.~J., Blake, J., Burgman, M.~A., Challenor, P., Coffeng,
  L.~E., Dawid, P., Angelis, D.~D., Goldstein, M., Hemming, V., Marion, G.,
  McKinley, T.~J., Overton, C.~E., Panovska-Griffiths, J., Pellis, L., Probert,
  W. J.~M., Shea, K., Villela, D. A.~M., and Vernon, I.~R. (2022).
\newblock Challenges in estimation, uncertainty quantification and elicitation
  for pandemic modelling.
\newblock {\em Epidemics}, 38:100547.

\bibitem[Thompson and Foege, 1968]{thompson1968}
Thompson, D. and Foege, W. (1968).
\newblock Faith tabernacle smallpox epidemic, {A}bakaliki, {N}igeria.
\newblock {\em World Health Organization}, WHO/SE:68.3.

\end{thebibliography}

\section*{Appendix}
Equation (\ref{Exactexp}) in the main paper is an exact expression of the Hellinger distance for a simplified case. Recall that the Hellinger distance can be used as an upper bound of the $\mathcal L_1$ distance between the target and proposal densities. A detailed proof of equation (\ref{Exactexp}) is included in this section. 

\subsection{Overview}\par
To compute the Hellinger distance given by equation (\ref{hellinger}) in the main paper, we need to do two integrals, namely $p(x)=\int f_Y(y|x)f_X(x|y)dy$ and $\tilde p(x)=\int f_Y(y|x)\sqrt{f_X(x|y)}dy$.
In this appendix we will focus on solving the integral $p(x)$ by computing the expectation of $f_X(x\mid Y)$ with respect to $Y\sim f_Y(y\mid x)$ and the procedure will be similar for $\tilde p(x)$.
Consider the bivariate, continuous-time point process $\{X_t,\ Y_t\}_{0\le t\le T}$ with infinitesimal probabilities
$$P(X_{t+h}=X_t+1\mid X_t,Y_t)=\beta(X_t,Y_t)h+o(h)$$
$$P(Y_{t+h}=Y_t+1\mid X_t,Y_t)=\gamma(X_t,Y_t)h +o(h),$$
where $\beta(X_t,Y_t)$ and $\gamma(X_t,Y_t)$ are the intensity functions, which only depend on the current state of the process. As defined in equation (\ref{path}) in the main paper, the right continuous sample paths $(x, y)$ can be characterized by the number and location of jumps, denoted by $(n_x, n_y)$, $\{t_i^x\}_{i=1}^{n_x}$ and $\{t_i^y\}_{i=1}^{n_y}$.
As explained in Section \ref{s:method2} of the main paper, the likelihood of $(X, Y)$ can be factorized as the product of likelihoods for two time-inhomogeneous birth processes, i.e. $p(x,\ y)=f_X(x\mid y)f_Y(y\mid x)$,
\begin{equation}
\label{fx}
   f_X(x\mid y) = \prod_{i=1}^{n_x}\ \beta(x(t_i^x-), y(t_i^x-))\ \text{exp}\left\{-\int_0^T\beta(x(t), y(t))dt\right\} 
\end{equation}
$$f_Y(y\mid x) = \prod_{i=1}^{n_y}\ \gamma(x(t_i^y-), y(t_i^y-))\ \text{exp}\left\{-\int_0^T\gamma(x(t), y(t))dt\right\}.$$
In the following part of the Appendix, we will consider $Y$ as the time-inhomogeneous birth process depend on $x$, with intensity function $\gamma(x(t),\ Y_t)$ and likelihood $f_Y(y|x)$ and consider $x$ as a fixed sample path. Therefore, back to the Hellinger distance, the two integrals can be expressed as the expectations with respect to $Y$,
$$p(x)=\text{E}_Y\left[f_X(x\mid Y)\right] \quad \mbox{and}\quad \tilde p(x)= \text{E}_Y\left[\sqrt{f_X(x\mid Y)}\right].$$

It is hard to find the above expectations directly since $Y$ is a continuous-time process. Therefore, we try to do a discretization for both $x$ and $Y$, and compute the expectation $\text{E}_{Y^n}[f_{X}(x_n\mid Y^n)]$ with a discrete-time count path $x_n$ and discrete-time count process $Y^n$. Then take the limit of the expectation as $n\rightarrow\infty$ to recover the desired expectations. The convergence of expectations can be shown by the convergence of $x_n\rightarrow x$ in Skorokhod topology and the convergence of $Y^n\rightarrow Y$ in distribution. In the following section, we will illustrate the convergence of expectations with the function $f_X(x\mid Y)$ and the proof is the same for $\sqrt{f_X(x\mid Y)}$.

\subsection{Discretization of the Process}\par
\label{discr}

Without loss of generality, instead of using the time interval $[0, T]$, we will consider $D[0, 1]$ \citep{billingsley2}, which is the space of real functions on $[0, 1]$ that are right-continuous with left-hand limits. Furthermore, let $D_c$ be the set of count paths in $D$. 

\subsubsection{Weak Convergence}
\label{weak}
Suppose $Y$ is a time-inhomogeneous birth process in $D_c$, with intensity function $\gamma(x,\ Y)$, where $x\in D_c$ is a fixed count path in $D_c$. Let $Y^n$ be a point process that can only jump at discrete time points $t \in T_n = \{0, 1/n, 2/n, \ldots, 1\}$, with probability $\gamma(x(t),\ Y_t)\frac1n$, so we have
$P(Y^n_{t+\frac1n} = Y^n_t+1) = \gamma(x(t),\ Y^n_t)\frac1n,\ \forall\ t\in T_n,\ t\ne1.$
Here we show that $Y^n\stackrel d\rightarrow Y$ by considering the convergence of the finite dimensional distributions; see \cite{billingsley2}.

Firstly, we will show that all the finite dimensional distributions of $Y_n$ converges weakly to the corresponding finite dimensional distributions of $Y$. For a given time point $t\in[0, 1]$, let $S_t$ and $S^n_t$ denote the waiting time until the next jump for $Y$ and $Y^n$, i.e.
$S_t = \text{inf}\{s\in[0, 1-t]: Y_{t+s} = Y_t + 1\}$ and $S^n_t = \text{inf}\{s\in[0, 1-t]: Y^n_{t+s} = Y^n_t + 1\}$.
It's sufficient to show, for any $t\in[0, 1]$, given $Y_t = Y^n_t = y(t)$, that $S^n_t$ converges in distribution to $S_t$, 
since both $Y$ and $Y^n$ are Markov processes.

The cumulative density functions are denoted by, $F_t(s) = P(S_t \le s\mid Y_t = y(t))$, and $F^n_t(s)= P(S^n_t \le s\mid Y^n_t=y(t))$, so we want to show that $F^n_t(s)\rightarrow F_t(s)$ for any $t\in[0, 1]$ and $s\in[0, 1-t]$.
For $Y$, the cumulative density function of waiting time is similar to the in-homogeneous Poisson process \citep{kingman1992}, 
$$F_t(s) = 1 - \exp\left\{ - \int_{t}^{t+s} \gamma\left(x(u), y(t)\right)du\right\}.$$
For $Y_n$, we can consider the probability that no jumps happened in $[t, t+s]$,
$$F^n_t(s)= 1 - \prod_{u\in T_n\cap[t, t+s]} \left\{1 - \gamma\left(x(u), y(t)\right)\frac1n\right\}.$$
As $n\rightarrow\infty$, using product integrals \citep{dollard2011}, the limit is given by
$\lim_{n\rightarrow\infty}F^n_t(s) = F_t(s).$
Therefore, conditional on the same value of current state at time $t$, the waiting time until the next jump of $Y^n$ converges in distribution to that of $Y$, i.e. $S^n_t \stackrel d\rightarrow S_t$ given $Y^n_t = Y_t$, which completes the proof. From \cite{billingsley2} Theorem 12.6, the convergence of finite dimensional distribution in $D_c$ implies weak convergence, so we have $Y^n\stackrel d\rightarrow Y$.

Now, consider $f_X(x\mid y)$ in equation (\ref{fx})
as a function of $y$ that maps the count path $y$ from $D_c$ to a real number. With any fixed path $x\in D_c$, this function is bounded since $n_x$ is finite and the  intensity function $\beta(x, y)$ is positive and bounded;
continuous almost everywhere with 
$\mu_Y(\{y:f_X\ \text{is discontinuous at}\ y\}) = 0$ for any $x\in D_c$
since the discontinuities happen only when $Y$ jumps at $t\in \{t_i^x\}_{i=1}^{n_x}$.

Now using the continuous mapping theorem (see Theorem 2.7 in \cite{billingsley2}), with $Y^n\stackrel d\rightarrow Y$, for the bounded and continuous a.e. function $f_X$,
we have the convergence of expectations, $\text{E}_{Y^n}[f_X(x\mid Y^n)] \rightarrow \text{E}_Y[f_X(x\mid Y)]$.

\subsubsection{Convergence in Skorohod Topology}
\label{skorohod}
Suppose $x$ is a count path in $D_c$ and
$x_n$ agrees with $x$ at discrete time points in $T_n := \{0, 1/n, 2/n, \ldots, 1\}$, and $x_n$ stays constant in each time interval $[(k-1)/n, k/n)$ for $k = 1,2,\ldots,n$, so $x_n$ is a right continuous count path in $D_c$, but can only jump at $t\in T_n$. Define the function $\lambda_n:[0, 1]\rightarrow T_n$ as a reformation of time such that,
$\lambda_n(t) := \text{max}\{s\in T_n: s\le t\}$, thus $x_n(t) = x(\lambda_n(t))\ \forall\ t\in [0, 1],$
with
$|\lambda_n(t) - t|<1/n $ for any $t\in[0, 1]$, so $\lambda_n(t)\rightarrow t$ uniformly and $x_n(t) \rightarrow x(t)$
for all the continuity points $t$ of $x$. Note that $x\in D_c$ only has finitely many discontinuities, so it follows $x_n\rightarrow x$ in Skorohod Topology \citep{billingsley2}.

Denote the jump times of $x$ and $x_n$ as $\{t_i^x\}_{i=1}^{n_x}$ and $\{t_i^{x_n}\}_{i=1}^{n_x}$. Then we have
$t_i^{x_n} = \min\{s\in T_n:s\ge t_i^x\}$ for $i=1, 2, \ldots, n_x,$
with
$x_n(t_i^{x_n}-) = x(t_i^x-) = i-1$ if $x_n(0) = x(0) = 0$. Note that two jumps can happen at the same time for $x_n$ when $n$ is small, but this is not a problem for large $n$ and as $n\rightarrow\infty$, since $T_n$ will become fine enough. 
With $|t_i^{x_n} - t_i^x|<1/n$, we have $t_i^{x_n}\rightarrow t_i^x$ and $y(t_i^{x_n}-)\rightarrow y(t_i^x-)$ for any $y\in D_c$ that are continuous at $\{t_i^x\}_{i=1}^{n_x}$.

Now, consider the likelihood function $f_X(x_n\mid y)$.
We have $x_n(t_i^{x_n}-) = x(t_i^x-)$, $x_n\rightarrow x$ in Skorohod topology, and $y(t_i^{x_n}-)\rightarrow y(t_i^x-)$
for any $y\in D_c$ that are continuous at $\{t_i^x\}_{i=1}^{n_x}$. For a continuous time count process $Y$, the probability of having a jump at any exact time point is 0, i.e. $\mu_Y(\{y: y\ \text{is discontinuous at} \{t_i^x\}_{i=1}^{n_x}\}) = 0$, so we can conclude that the convergence of $x_n\rightarrow x$ in Skorohod topology leads to the convergence of expectations given by,
$\text{E}_{Y}[f_{X}(x_n\mid Y)]\rightarrow \text{E}_{Y}[f_X(x\mid Y)].$

\subsubsection{Convergence of Expectations}

We want to show the convergence of expectations $\text{E}_{Y^n}[f_{X}(x_n\mid Y^n)] \rightarrow \text{E}_Y[f_X(x\mid Y)]$. 
By the triangle inequality,
\begin{align*}
 & |\text{E}_{Y^n}[f_{X}(x_n\mid Y^n)]-\text{E}_Y[f_X(x\mid Y)]|
 \\
 \le\  & |\text{E}_{Y^n}[f_{X}(x_n\mid Y^n)]-\text{E}_{Y}[f_X(x_n\mid Y)]|\ + 
 |\text{E}_{Y}[f_X(x_n\mid Y)]-\text{E}_Y[f_X(x\mid Y)]|.   
\end{align*}
From Section \ref{weak}, we have $\text{E}_{Y^n}[f_X(x\mid Y^n)] \rightarrow \text{E}_Y[f_X(x\mid Y)]$. Note that, the distribution of $Y^n$ with $x_n$ is exactly the same as that with $x$, since $x_n(t) = x(t)$ for any $t\in T_n$. Therefore, $(Y^n\mid x) \stackrel d\rightarrow (Y\mid x)$ is equivalent to $(Y^n\mid x_n)\stackrel d\rightarrow (Y\mid x)$ and it follows $\text{E}_{Y^n}[f_X(x_n\mid Y^n)] \rightarrow \text{E}_Y[f_X(x_n\mid Y)]$.
From Section \ref{skorohod}, we have $\text{E}_{Y}[f_{X}(x_n\mid Y)]\rightarrow \text{E}_{Y}[f_X(x\mid Y)]$.
Therefore, by putting together section \ref{weak} and \ref{skorohod}, we have
$\text{E}_{Y^n}[f_{X}(x_n\mid Y^n)] \rightarrow \text{E}_Y[f_X(x\mid Y)]$.

\subsection{Computing the Expectations}

Here we compute the expectation $\text{E}_{Y^n}[f_{X}(x_n\mid Y^n)]$ by using the discretization in section \ref{discr} and then take the limit for $n\rightarrow\infty$ to get the desired expectation $\text{E}_{Y}[f_{X}(x\mid Y)]$. For notation simplification, in the following computation, we will write $x_n(k/n)$ and $y_n(k/n)$ as $x(k)$ and $y(k)$, and let $h = 1/n$.

Let $Y^n$ be the discrete-time point process as defined in Section \ref{weak}, with probability function given by
$P_{Y^n}(y_n\mid x_n) = \prod_{k=0}^{n-1} P(y(k+1)\mid x(k),y(k))$,
where $P\left(y(k+1)\mid x(k),y(k)\right) = \gamma\left(x(k), y(k)\right)h$ for $y(k+1) = y(k)+1$, and $P(y(k+1)\ |\ x(k),y(k)) = 1-\gamma(x(k), y(k))h$ for $y(k+1) = y(k)$.
The function $f_X(x_n\mid y_n)$ can be written in a similar product form, 
$$f_X(x_n\mid y_n) =\prod_{k=0}^{n-1} f(x(k+1)\mid x(k),y(k)),$$ where 
$f(x(k+1)|x(k),y(k)) = \beta(x(k), y(k))\exp\left\{-\beta(x(k), y(k))h\right\}$ for $x(k+1) = x(k)+1$ and $f(x(k+1)|x(k),y(k)) = \exp\left\{-\beta(x(k), y(k)\big)h\right\}$ for $x(k+1) = x(k)$.
Let $\mathcal{Y}^n$ denote the support set of $Y^n$. We can compute $\text{E}_{Y^n}[f_{X}(x_n\mid Y^n)]$ by summing over all $y_n\in \mathcal{Y}^n$.   
\begin{align*}
     \text{E}_{Y^n}[f_{X}(x_n\mid Y^n)]
    = & \sum_{y \in \mathcal{Y}^n}\ \prod_{k=0}^{n-1} f(x(k+1)| x(k),y(k))\ \prod_{k=0}^{n-1} P(y(k+1)|x(k),y(k)) \\
    = & f(x(1)| x(0),y(0))  \sum_{y(1)}
    f(x(2)| x(1),y(1))P(y(1)| x(0),y(0)) \ldots \\
     & \sum_{y(n-1)} f(x(n)| x(n-1),y(n-1))P(y(n-1)| x(n-2),y(n-2)).
\end{align*}
To further simplify the notations, define
$$w_k := f(x(k+1)\mid x(k),y(k))\ P(y(k)\mid x(k-1),y(k-1)).$$
We can first compute $\sum_{y(n-1)}w_{n-1}$, and then $\sum_{y(n-2)}w_{n-2}\sum_{y(n-1)}w_{n-1}$, etc, so that
$$\text{E}_{Y^n}[f_{X}(x_n\mid Y^n)]=f(x(1)\mid x(0),y(0))\sum_{y(1)}w_1\ldots\sum_{y(n-1)}w_{n-1},$$
can be computed recursively from the end towards the beginning.

Here we derive $\text{E}_{Y^n}[f_{X}(x_n\mid Y^n)]$, by recursion, for a special case. Assume $x(0) = y(0) = 0$, $\ \gamma(x,y)=\gamma\ $ and $\ \beta(x,y)=\beta_0+\epsilon y\ $, so $\beta(x, y+1) - \beta(x,y)=\epsilon$ and we assume $x$ has no jumps in the time interval $[0, 1]$, so $x(k+1) = x(k)$ for each $k$, and we have
$$w_k = \text{exp}[-\{\beta_0 + \epsilon\ y(k)\}h] \ P(y(k)\mid x(k-1),y(k-1)).$$
The recursion starts from $$\sum_{y(n-1)}w_{n-1}=\text{exp}\left[-\{\beta_0 + \epsilon\ y(n-2)\}\ h\right] \ \{1 - \gamma ( 1 - e^{-\epsilon h}) h\}.$$
The next step is given by,
\begin{align*}
\sum_{y(n-2)}w_{n-2}\sum_{y(n-1)}w_{n-1} =& \{1 - \gamma (1 - e^{-\epsilon h}) h\} \{1 - \gamma (1 - e^{-2\epsilon h}) h\} \text{exp}\left[-2\{\beta_0 + \epsilon\ y(n-3)\}h\right].
\end{align*}
Keep repeating this, to observe the pattern.
At the end of the recursion we will get
$$\text{E}_{Y^n}[f_{X}(x_n\mid Y^n)]=\text{exp}\left[-n\left\{\beta_0 + \epsilon\ y(0)\right\}\ h\right] \prod_{k=1}^{n-1} \left\{1 - \gamma ( 1 - e^{-k\epsilon h}) h\right\}.$$
Now take the limit as $n\rightarrow\infty$ and by product integral,
\begin{align*}
   \text{E}_{Y}\left[f_{X}(x\mid Y)\right] =\lim_{n\rightarrow\infty}\text{E}_{Y^n}\left[f_{X}(x_n\mid Y^n)\right] 
    =  \exp\left\{-\beta_0 - \int_0^1 \gamma \left( 1 - e^{-\epsilon t}\right) dt \right\}.
\end{align*}
Clearly, $\text{E}_{Y}\left[\sqrt{f_{X}(x\mid Y)}\right]$ can be derived in a similar way, simply by replacing all the $\text{exp}\left[-\left\{\beta_0 + \epsilon\ y(k)\right\}h\right]$ 
with $\text{exp}\left[-\frac12\{\beta_0 + \epsilon\ y(k)\}h\right]$, so
$$\text{E}_{Y}\left[\sqrt{f_{X}(x\mid Y)}\right]=\exp\left\{-\frac12\beta_0 - \int_0^1 \gamma \left( 1 - e^{-\frac12\epsilon t}\right) dt \right\}.$$
Finally, we change the time interval $[0, 1]$ back to $[0, T]$ and get equation (\ref{Exactexp}), i.e.
\begin{align*}
    H^2\left(p(y'\mid x)f_Y(y\mid x),\ p(y\mid x)f_Y(y'\mid x)\right) 
    =1-\exp\left[-\gamma\left\{T-\epsilon^{-1}\left(e^{-\epsilon T}-4e^{-\frac12\epsilon T}+3\right)\right\}\right].
\end{align*}

\label{lastpage}

\end{document}